\documentclass[aps,prd,11pt, tightenlines, twoside, secnumarabic, 
superscriptaddress, showpacs, preprintnumbers, nofootinbib, notitlepage,
onecolumn, fleqn]{revtex4-1}

\pdfoutput=1
\usepackage[english]{babel}
\usepackage{amsmath,amssymb,amsfonts, bm,bbm,slashed}
\usepackage{graphicx}
\usepackage[sort&compress]{natbib}
\usepackage{xcolor}
\usepackage[normalem]{ulem}
\usepackage{hyperref}
\usepackage{cleveref}
\definecolor{red}{rgb}{1.0, 0, 0}
\usepackage{hyperref}
\usepackage{enumerate}
\usepackage{epsfig, subfigure}
\usepackage{setspace}
\usepackage{booktabs, tabularx}
\usepackage{units}

\hypersetup{
    pdfnewwindow=true,      % links in new window
    colorlinks=true,       % false: boxed links; true: colored links
    linkcolor=blue,          % color of internal links
    citecolor=blue,        % color of links to bibliography
    filecolor=blue,      % color of file links
    urlcolor=blue        % color of external links
}

\allowdisplaybreaks

\bibpunct{[}{]}{,}{n}{}{,}

\newcommand{\ev}[1]{\ensuremath{\left\langle #1 %
                     \right\rangle}} % Expectation value
\begin{document}

%--------------------------------------------------------------------%
\title{Photons, Photon Jets and Dark Photons at 750\,GeV and Beyond}

\author{Basudeb Dasgupta}
\email{bdasgupta@theory.tifr.res.in}
\affiliation{Tata Institute of Fundamental Research,
             Homi Bhabha Road, Mumbai, 400005, India.}

\author{Joachim Kopp}
\email{jkopp@mpi-hd.mpg.de}
\affiliation{PRISMA Cluster of Excellence \& Mainz Institute for Theoretical Physics,
             Johannes Gutenberg University, Staudingerweg 7, 55099 Mainz, Germany}

\author{Pedro Schwaller}
\email{pedro.schwaller@desy.de}
\affiliation{DESY, Notkestrasse 85, D-22607 Hamburg, Germany}
%\affiliation{CERN, Theory Division, CH-1211 Geneva, Switzerland}

\date{\today}
\pacs{}
\preprint{TIFR/TH/16-05, MITP/16-020, DESY 16-028}
%--------------------------------------------------------------------%

\begin{abstract}
  In new physics searches involving photons at the LHC, one
  challenge is to distinguish scenarios with isolated photons from models
  leading to ``photon jets''. For instance, in the context of the 750~GeV
  diphoton excess, it was pointed out that a true diphoton resonance $S \to
  \gamma\gamma$ can be mimicked by a process of the form $p p \to S \to a a \to
  4\gamma$, where $S$ is a new scalar with a mass of 750~GeV and $a$ is a light
  pseudoscalar decaying to two collinear photons.  Photon jets can be
  distinguished from isolated photons by exploiting the fact that a large fraction of
  photons convert to an $e^+e^-$ pair inside the inner detector.  In this
  note, we quantify this discrimination power, and we study how the
  sensitivity of future searches differs for photon jets compared to isolated
  photons. We also investigate how our results depend on the lifetime of
  the particle(s) decaying to the photon jet. Finally, we discuss the extension 
  to $S\to A^\prime A^\prime\to e^+e^-e^+e^-$, where there are no photons at all 
  but the dark photon $A^\prime$ decays to $e^+e^-$ pairs. Our results will be useful in
  future studies of the putative 750~GeV signal, but also more generally in any
  new physics search involving hard photons.
\end{abstract}

%--------------------------------------------------------------------%
\maketitle
%--------------------------------------------------------------------%

%--------------------------------------------------------------------%
\section{Introduction}
\label{sec:intro}
%--------------------------------------------------------------------%

In their recent end-of-year jamboree, the ATLAS and CMS collaborations
have reported an impressive cornucopia of LHC Run~II results. One of them---a
possible excess in the two photon final state at an invariant mass
$\sim 750$~GeV~\cite{ATLAS-CONF-2015-081, CMS-PAS-EXO-15-004}---has
caused a flurry of discussion in the
community~\cite{DiphotonInspire}.
Most of these works introduce
a new neutral scalar particle $\phi$ with a mass around 750~GeV and decaying to
two photons. Both the production and the decay of this particle typically proceed
through loop diagrams. Constraints from Run~I data imply that
the production cross section of $\phi$ must be significantly
larger at the Run~II center-of-mass energy of 13~TeV than at the Run~I
energy of 8~TeV. Moreover, decay modes of $\phi$ other than $\phi \to \gamma\gamma$
should not be too strong, but at the same time, $\phi$ should have a large
total width $\sim 45$~GeV to optimally fit the data.

A very appealing class of alternative models explaining the 750~GeV excess
are those in which the final state is in fact not two body, but contains
two ``photon jets'', i.e.\ groups of highly collinear photons \cite{Draper:2012xt,Ellis:2012zp,Ellis:2012sd,Dobrescu:2000jt,
Toro:2012sv,Chang:2006bw,Curtin:2013fra}. 
If the photon jets are sufficiently collimated, they are indistinguishable from isolated photons using information from the electromagnetic calorimeter alone. Therefore, models of this type could explain the diphoton anomaly, as discussed in Refs.~\cite{Knapen:2015dap,Agrawal:2015dbf,Chala:2015cev,Aparicio:2016iwr,Chang:2015sdy,Ellwanger:2016qax}. While the experiments have strong discriminating variables to reject e.g.\ photon pairs coming from neutral hadron decays, the studies~\cite{Knapen:2015dap,Agrawal:2015dbf,Chala:2015cev,Aparicio:2016iwr,Chang:2015sdy,Ellwanger:2016qax} show that there are regions of parameter space where the photon jets are expected to pass the tight photon selection. 

\begin{figure}
\includegraphics[width=0.6\textwidth]{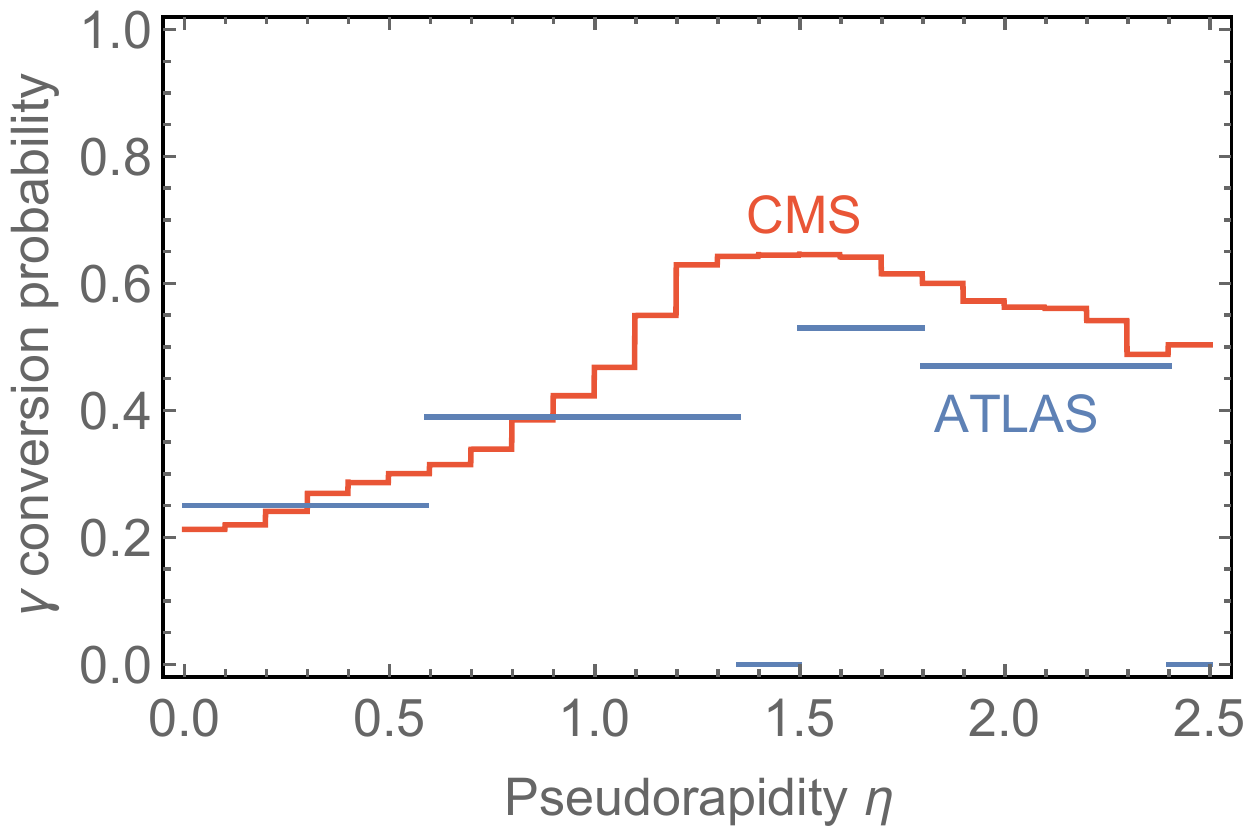}
\caption{Probability for prompt photons to be reconstructed as converted photons in the ATLAS (blue) and CMS (red) detectors, based on 13~TeV data (ATLAS, Ref.~\cite{ATLAS-gamma-conv}) and 8~TeV data (CMS, Ref.~\cite{Khachatryan:2015iwa}). 
}
\label{fig:conversiondata}
\end{figure}

However, there is a catch: since photons have to travel through some
amount of detector material before reaching the calorimeter, they have
a high 
(e.g.~$\sim 40\%$ at ATLAS~\cite{ATLAS-gamma-conv}) 
probability of converting to
an $e^+ e^-$ pair already in the inner detector, 
with nontrivial pseudorapidity dependence (see Fig.~\ref{fig:conversiondata}). 
Such conversions can occur in
the strong electric field of an atomic nucleus through a process $\gamma + Z
\to Z + e^+ + e^-$.  ``Converted photons''
are routinely included in analyses involving photons.

For a high-$p_T$ photon jet with $\geq 2$ photons, it is clear that the probability that at least one of the photons inside the jet converts is higher than for isolated photons. Even if no further discrimination is performed, we will show below that the ratio of converted to unconverted photon events already provides a powerful discrimination between isolated photon and photon jet models. Furthermore this ratio can also be used to improve the sensitivity of searches for photon jet events. 

Going beyond  conversion ratios, several other observables could be used to reveal the photon jet origin of signals involving photons, including non-resonant photons. This includes a mismatch of the track $p_T$ and the calorimeter $E_T$ if only one photon converts, a non-standard response of the signal to changes of the photon selection criteria, and converted photon candidates with more than two tracks. 

In the rest of this note, we will first discuss the use of converted photon
ratios to discriminate events with photon jets from isolated photons and to
improve the sensitivity of searches for such models (\cref{sec:photon-jets} and
\cref{sec:converted-gammas}). After that we will analyze the effects of finite
lifetime of the intermediate states on this analysis (\cref{sec:finite-tau}),
and we will extend the discussion to models with dark photons decaying directly
to displaced $e^+ e^-$ pairs (\cref{sec:dark photons}). Finally, we discuss
the prospects of other observables in more detail
(\cref{sec:other-observables}). 
While most of our numerical results are obtained using ATLAS 13~TeV data, we expect that at CMS similar results can be expected, since the conversion rate is similar in magnitude and rapidity dependence, as seen from Fig.~\ref{fig:conversiondata}.

%--------------------------------------------------------------------%
\section{Photon Jets}
\label{sec:photon-jets}
%--------------------------------------------------------------------%

Before digging into the details, let us first review the type of models that
can give rise to photon jets and which therefore can be probed by the methods
we present below. Any particle that decays to two or more photons can produce a
photon jet if it is sufficiently boosted. Consider a particle $a$ with mass $m_a$ and
Lorentz boost $\gamma = E /m_a$ decaying to two photons. The 
minimal opening angle between the two photons is
\begin{align}
	\Delta \phi_{\rm min} & = \arccos \left( 1- \frac{2}{\gamma^2} \right) \approx \frac{2}{\gamma} \,.
\end{align}
Experimentally, photon pairs with opening angles below $\Delta R \sim 0.01$ are
difficult to distinguish from isolated photons in the calorimeter. Therefore if
$a$ is produced in the decay of a TeV scale resonance, one finds that for $m_a
\lesssim 2$~GeV the photon pairs from each $a$ decay can easily pass as
isolated photon candidates\footnote{To be more precise, the first layer of the
EM calorimeter in ATLAS is very finely segmented with $\Delta \eta \approx
0.002-0.003$, and shower shape variables are used to suppress backgrounds from
$\pi_0$ decays. So the actual bound on $m_a$ could be as low as 500~MeV, as
argued in~\cite{Knapen:2015dap}. The main point here is that there is a region
of parameter space where collimated photon jets can pass as single isolated
photons, so the precise value of the limit is not important.}. Models of this
type were considered before in the context of exotic Higgs
decays\cite{Chang:2006bw,Draper:2012xt,Curtin:2013fra} and more recently as
alternative interpretations of the 750~GeV
resonance~\cite{Knapen:2015dap,Agrawal:2015dbf,Chala:2015cev,Aparicio:2016iwr,Chang:2015sdy,Ellwanger:2016qax}. 

Couplings of a light state $a$ to photons are also constrained by low
energy data~\cite{Bjorken:2009mm,Jaeckel:2010ni,Essig:2010gu,Essig:2013lka,
Alekhin:2015byh,Jaeckel:2015jla,Dobrich:2015jyk}.
This makes it impossible to choose $m_a$ arbitrarily small.
Nevertheless, $m_a$ could be so small that its (laboratory frame)
decay length becomes comparable to or even larger than the size of the
ATLAS and CMS inner detectors (about a meter).  If $a$ decays to $\gamma\gamma$
at a macroscopic distance from the beam pipe, but still within the inner
detector, the two photons have a smaller conversion probability than for
quasi-instantaneous $a$ decay. We will consider this possibility in \cref{sec:finite-tau}.
If the decay length of $a$ is so large,
that most decays occur outside the electromagnetic calorimeter, they
can no longer mimic isolated photons. Such
scenarios are, however, still of phenomenological interest in the context of
displaced object searches, which look for objects decaying in the
calorimeters or in the muon system~\cite{Aad:2015asa,Aad:2015rba}.

\begin{figure}
\includegraphics[width=0.5\textwidth]{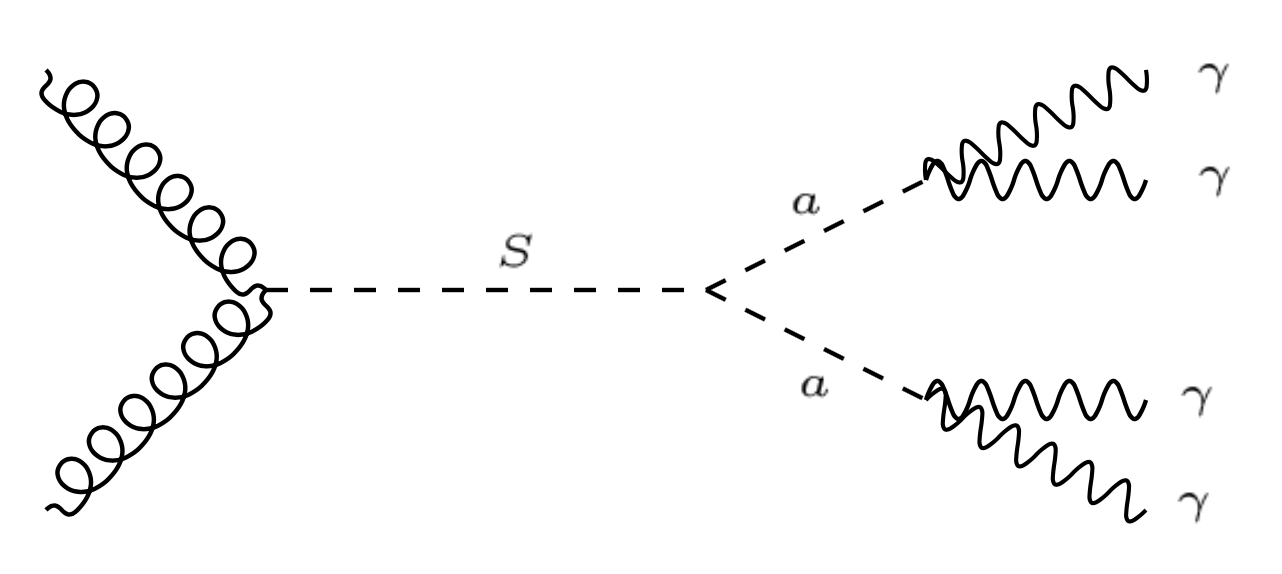}
\caption{Feynman diagram illustrating the production of a scalar resonance $S$ followed by the decay into collimated photon jets.}
\label{fig:feynman}
\end{figure}

To be as model independent as possible, we will consider scenarios where a resonance $X$ is produced in proton-proton collisions and decays to two light particles $a_1$, $a_2$, each of which in turn decays to $N_i$ photons: 
\begin{align}
  p p \to X \to (a_1 \to N_1 \gamma) + (a_2 \to N_2 \gamma) \,.
\end{align}
As a concrete realization, consider the case of a scalar resonance $S$ with loop induced couplings to gluons and tree level couplings to a light pseudo-scalar $a$, which in turn couples to photons: 
\begin{align}
	{\cal L} \supset -M_S^2 S^2 - m_a^2 a^2 + \frac{1}{\Lambda} S G_{\mu\nu} G^{\mu \nu} + \lambda S a a + \frac{1}{f} a F_{\mu\nu}{\tilde F}^{\mu\nu}\,.
        \label{eq:L1}
\end{align}
Here, $m_S$ and $m_a$ are the masses of scalar and pseudoscalar, respectively, and $1/\Lambda$,
$1/f$, $\lambda$ are coupling constants.
An LHC process which is induced by these couplings is shown in Fig.~\ref{fig:feynman}.
The five dimensionful parameters in \cref{eq:L1} are a priori independent and can be extracted from the data. The position of the peak in the photon invariant mass peak determines $M_S$, and the signal cross section together with the decay width of $S$ determines $\Lambda$ and $\lambda$.
$m_a$ and $f$ have to be chosen such that the photon jets pass as regular photons, which is non-trivial since the coupling $f$ of a light pseudo-scalar to photons is strongly constrained~\cite{Chala:2015cev}. 

%--------------------------------------------------------------------%
\section{Distinguishing Photon Jets from Isolated Photons}
\label{sec:converted-gammas}
%--------------------------------------------------------------------%
Consider a photon jet consisting of $N$ collimated photons. A regular isolated
photon corresponds to $N=1$ in this notation. A photon
jet will be registered as a converted photon if at least one of the photons
inside the jet converts and leaves a signal in the tracker. For a given
conversion rate $p^\text{conv}$ for individual photons in a given jet, the probability
that the photon jet appears as a converted photon is then given by
\begin{align}
  p_{N}^\text{conv} &= 1- \big[ 1 - p^\text{conv} \big]^N\,.
\label{eq:pN}
\end{align}
Obviously, the probability that the photon jet appears as an unconverted event is
\begin{align}
p_{N}^\text{no-conv} &= 1-p_{N}^\text{conv}\,.
\label{pNnc}
\end{align}
For the moment, we neglect the possible issue arising from having more than two
reconstructed tracks associated with the photon candidate, which could make the
photon fail isolation criteria. We will come back to this point later. 

Now consider a diphoton event\footnote{More precisely, an event with two
reconstructed photon candidates which are well separated from each other.}
with angular separation $\Delta R > 0.4$ so that
they do not overlap.  Microscopically, the event contains two photon jets,
with the number of photons in them denoted by $N_1$ and $N_2$.  From the
experimental point of view, we distinguish three event categories, namely
events with $i=0,1,2$ of the photon jets being reconstructed as
converted photons.  The probabilities $p^{(i)}_{(N_1 N_2)}$
for an event to fall into each of these categories depend on $N_1$ and $N_2$,
and thus offer a handle for distinguishing different theoretical models underlying
a diphoton signal. It is easy to see that
\begin{align}
  p^{(0)}_{(N_1 N_2)}
   &= p_{N_1}^\text{no-conv} \, p_{N_2}^\text{no-conv} \,,
                                  \label{eq:p0} \\
  p^{(1)}_{(N_1 N_2)}
   &= p_{N_1}^\text{no-conv} \, p_{N_2}^\text{conv}
    + p_{N_1}^\text{conv} \, p_{N_2}^\text{no-conv} \,,
                                  \label{eq:p1} \\
  p^{(2)}_{(N_1 N_2)}
   &= p_{N_1}^\text{conv} \, p_{N_2}^\text{conv} \,.
                                  \label{eq:p2}
\end{align}
In the following, we will in particular consider the prospects for distinguishing
a real diphoton resonance, $(N_1  N_2) = (1 1)$, from models with
$(N_1  N_2) = (1 2)$ or $(2 2)$, which have been proposed in the literature
as alternative explanations of the 750~GeV signal~\cite{Knapen:2015dap,Agrawal:2015dbf,Chala:2015cev,Aparicio:2016iwr,Chang:2015sdy,Ellwanger:2016qax} .
This discrimination is complicated by the fact that there is a
significant number of SM background events in the signal region. Here we assume that all
background events are of $(1 1)$ type but we expect the results to remain similar for any other known background composition (See \cref{sec:backgrnds} for details).

Perhaps the simplest statistical way of approximately quantifying the model
discrimination power is a Pearson $\chi^2$ test based on the following $\chi^2$
function:
\begin{align}
  \chi^2(S,B)
    &= \frac{S^2}{B} \sum_{i=0}^2 
    \frac{\Big(\, p^{(i)}_{(N_1^\text{true} N_2^\text{true})}
          -  \, p^{(i)}_{(N_1^\text{test} N_2^\text{test})} \Big)^2}
         { \frac{S}{B} \,  p^{(i)}_{(N_1^\text{test} N_2^\text{test})}
          + \, p^{(i)}_{(11)}} \,.
  \label{eq:chi2-eta}
\end{align}
Here, $S$ and $B$ are the numbers of signal and background events, respectively,
$(N_1^\text{true} N_2^\text{true})$ corresponds to the model we assume to be realized
in nature, while $(N_1^\text{test} N_2^\text{test})$ describes the model we wish
to test against.
In other words, the question we are asking here is how likely it is that the
hypothesis $(N_1^\text{test} N_2^\text{test})$ is accepted if the actual events
are of type $(N_1^\text{true} N_2^\text{true})$. Obviously, the right hand side of
\cref{eq:chi2-eta}
vanishes if $(N_1^\text{test} N_2^\text{test}) = (N_1^\text{true} N_2^\text{true})$.

The two jets have different $p_T$ and pseudorapidities $\eta$, which lead to 
unequal $p^\text{conv}$ for the photons in different jets. To account for this, we 
take the $p_T$- and $\eta$-dependent conversion probabilities $p^\text{conv}(p_T, \eta)$ 
given in ref.~\cite{ATLAS-gamma-conv}.
As the $p_T$-dependence of $p^\text{conv}(p_T, \eta)$ is weak for photons above
100~GeV, we neglect it in the following and work with $p^\text{conv}(\eta)$
depending only on the pseudorapidity. The value of $p^\text{conv}(\eta)$
in each $\eta$ bin is listed in \cref{tab:eta-binning} in \cref{sec:statistics}.
The $\chi^2$ function in \cref{eq:chi2-eta} is generalized
to also include a sum over the events in different bins $(jk)$ labeled by the pseudorapidities $(\eta_j,\eta_k)$ of the two jets. The probabilities 
$p^{(i)}_{(N_1 N_2)}(\eta_j, \eta_k)$ for $i$ conversions in
an event in rapidity bin $(jk)$ are given by \cref{eq:p0,eq:p1,eq:p2} using the appropriate $p^\text{conv}_{N_1}({\eta_1})$ and $p^\text{conv}_{N_2}({\eta_2})$ for each jet. Additionally, both terms in the numerator of \cref{eq:chi2-eta} as well as the first term in the denominator must now be multiplied by $p_S^{jk}$, the probability in the respective true/test model for signal events to fall into that rapidity bin. Similarly the second term in the denominator must now be multiplied by the analogous probability $p_B^{jk}$ for background events. These probabilities $p_S^{jk}$ and $p_B^{jk}$ can be obtained
by computing the differential cross sections for the signal and background
$(N_1 N_2)$.  We do so using MadGraph~5 v2.3.3~\cite{Alwall:2011uj,Alwall:2014hca},
 with a FeynRules / UFO~\cite{Alloul:2013bka}
implementation of a simple $(N_1 N_2) = (11)$ model that augments the Standard Model with
a scalar $S$ and the effective couplings
\begin{align}
  \mathcal{L} \supset \frac{1}{\Lambda_g} \, S \, G_{\mu\nu} G^{\mu\nu}
                    + \frac{1}{\Lambda_\gamma} \, S \, F_{\mu\nu} F^{\mu\nu} \,.
  \label{eq:L2}
\end{align}
Note that binning the data in pseudo-rapidity $\eta$ introduces some model
dependence since the differential rapidity distribution will be different from
model to model.  We assume in the following that the $\eta$ distribution of the
photon jets in models with $N_1, N_2 > 1$ is identical to the $\eta$
distribution of the isolated photons following from \cref{eq:L2}.  

In our numerical results, we will go somewhat beyond the $\chi^2$ test based
on \cref{eq:chi2-eta}, and instead employ a slightly more sensitive likelihood
ratio test, as discussed in \cref{sec:statistics}.  

\begin{figure}
  \begin{center}
    \begin{tabular}{ccc}
      \includegraphics[width=0.33\textwidth]{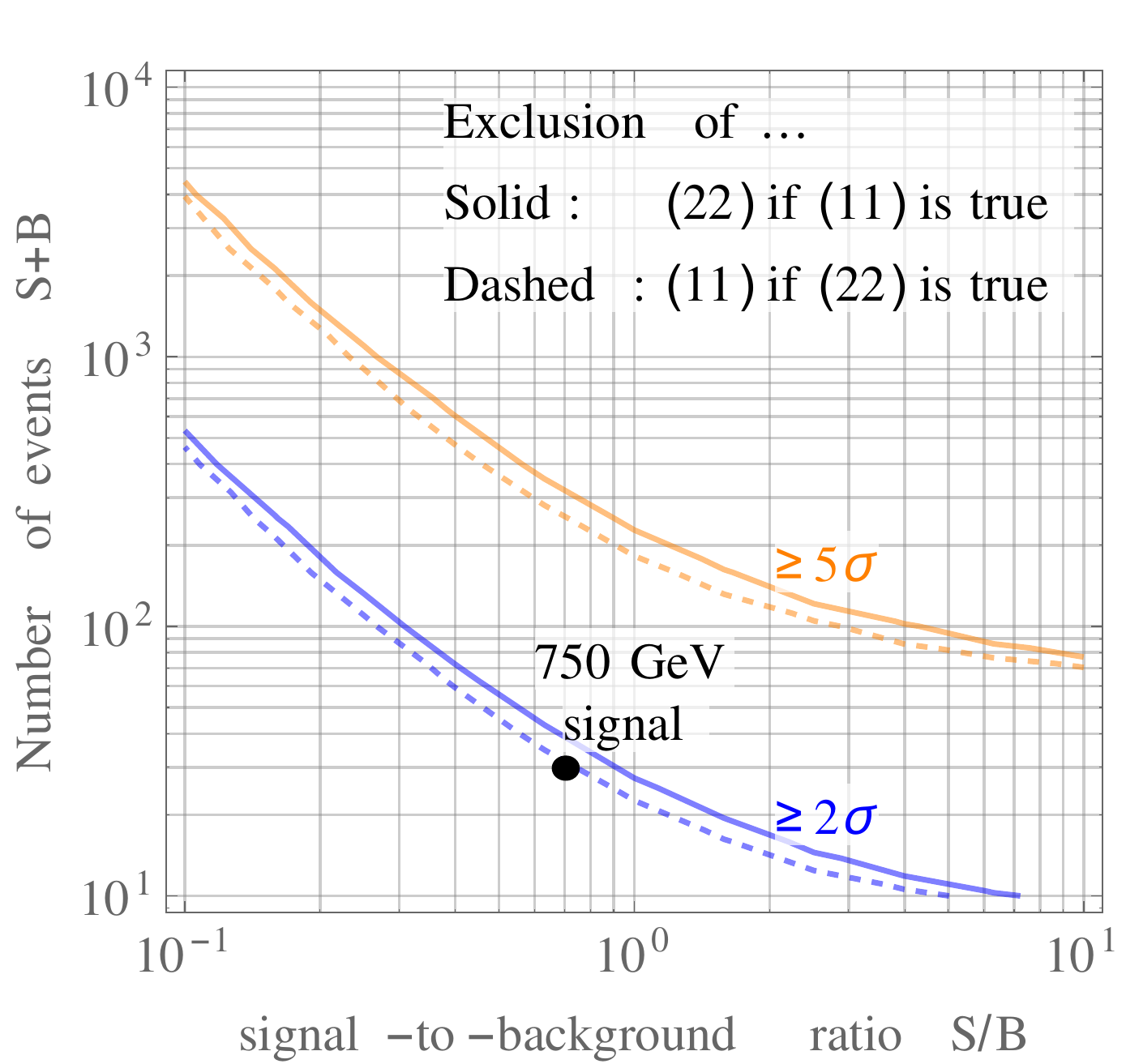} &
      \includegraphics[width=0.33\textwidth]{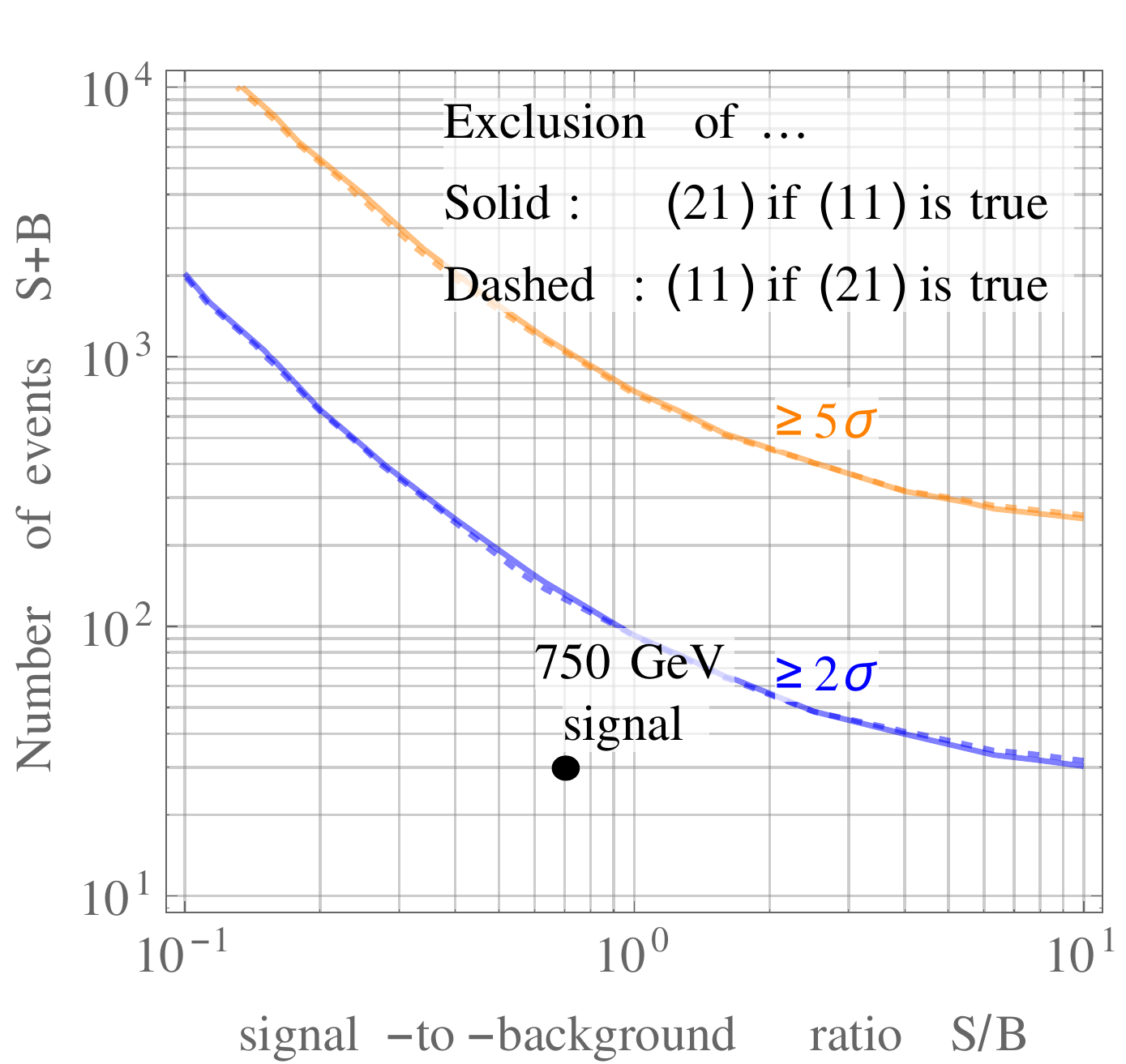} &
      \includegraphics[width=0.33\textwidth]{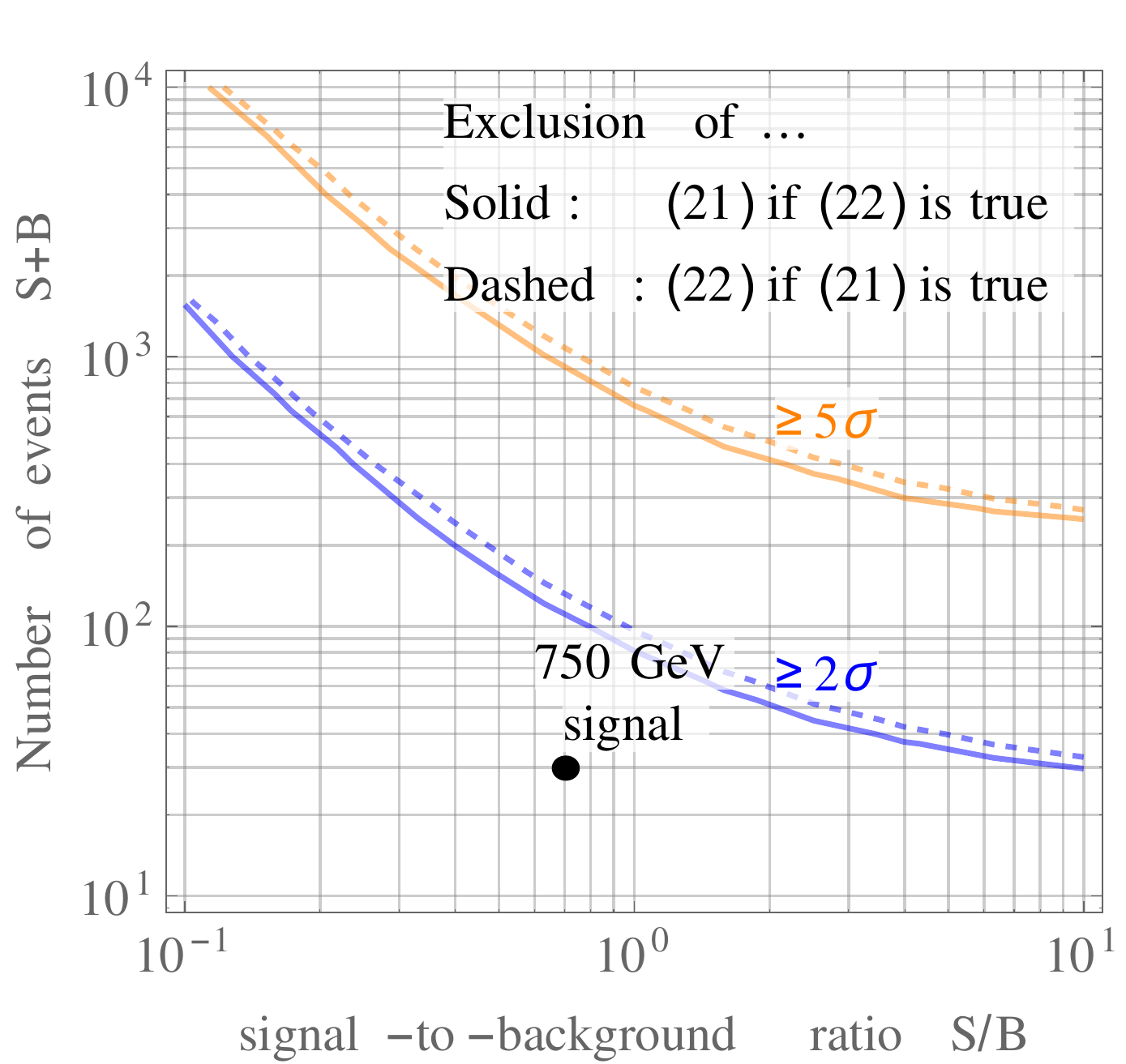} \\
      (a) & (b) & (c)
    \end{tabular}
  \end{center}
  \vspace{-0.5cm}
  \caption{The expected number of diphoton events required to discriminate between
    different model hypotheses $(N_1 N_2)$ based on different conversion rates.
    Here, $N_1$ and $N_2$ are the multiplicities of the two photons or photon
    jets in the event.  Results are shown as a function of the expected signal
    to background ratio.  The sensitivities shown here are based on the
    likelihood ratio test discussed in \cref{sec:statistics} and include the
    $\eta$-dependence of the photon conversion probability,}
  \label{fig:sigSB}
\end{figure}

For a given $S/B$, we can now ask how many expected events $S+B$ in the signal
region are needed to reject different hypotheses $(N_1 N_2)$ at the $2\sigma$
and $5\sigma$ level.  The results are shown in Fig.~\ref{fig:sigSB}.  For $S/B$
of order one, we see that at most a hundred events are necessary to distinguish
the different hypotheses at the $2\sigma$ level.  Discrimination between models
of type $(11)$ and $(22)$ model requires fewer events than discrimination
between the $(21)$ and $(11)$ or between the $(21)$ and $(22)$ scenarios. The
reason is simply that the conversion probabilities \cref{eq:p0,eq:p1,eq:p2} for
the two alternative hypotheses are more different in the former case.  For the
particular case of the excess observed around 750~GeV, the present data could
already be sufficient to discriminated between the $(11)$ and $(22)$ hypotheses
at the $2\sigma$ level, while more data would be needed to tell the $(21)$
hypothesis apart from either $(11)$ or $(22)$ scenarios.

We see that photon conversion rates are a promising tool to distinguish
between different new physics models in diphoton events once a signal is
observed.  However, also without an observed event excess, the different
conversion probabilities for isolated photons and $N>1$ photon jets can be
employed as an additional tool to discriminate photon jet signals from the
background. In the following we illustrate this, again using the example of a
search for a diphoton resonance in the mass range between 200~GeV and 1500~GeV. 

\begin{figure}
  \begin{center}
    \includegraphics[width=11cm]{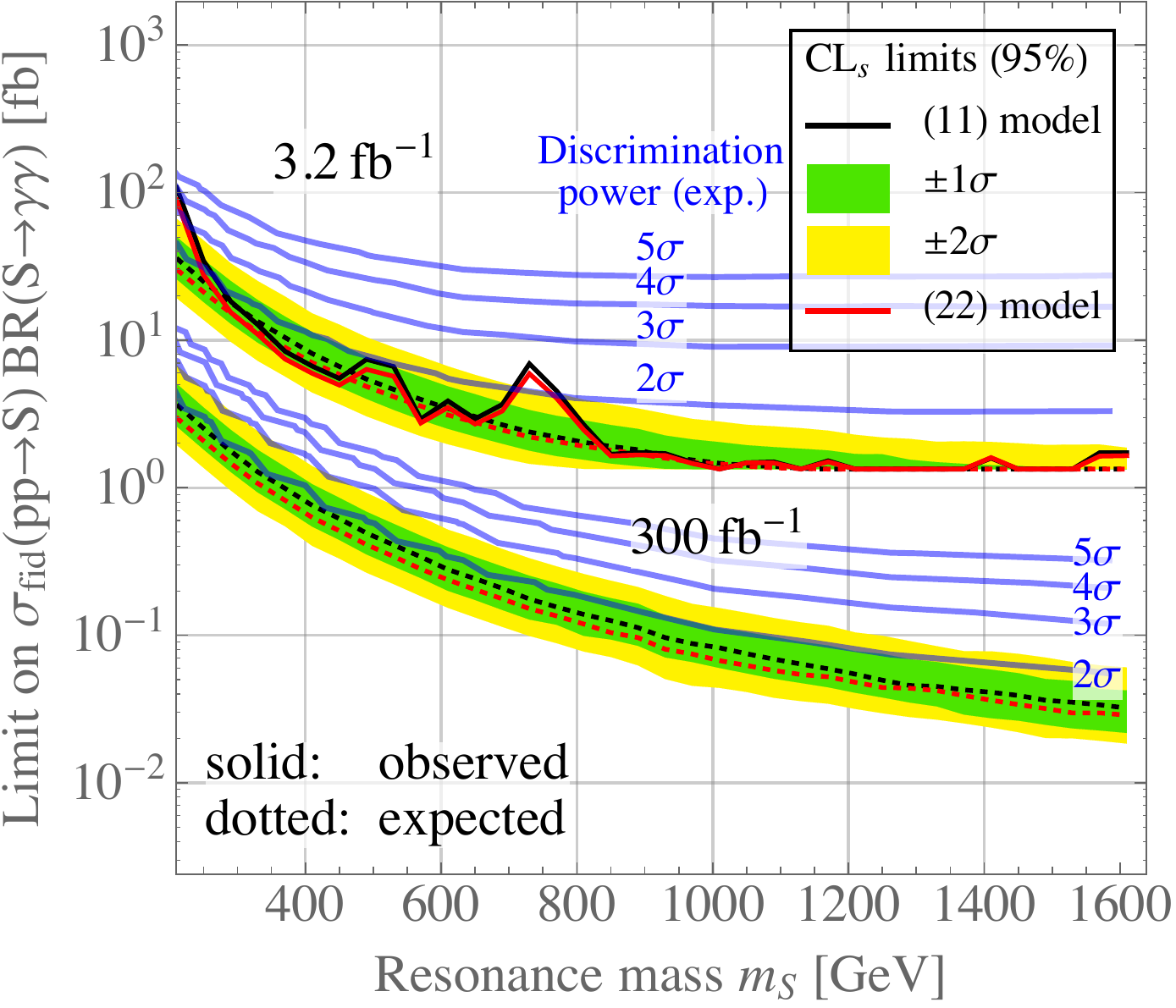}
  \end{center}
  \caption{The expected (dashed) and observed (solid) 95\% CL$_s$ limits on a
    true diphoton signal (the $(11)$ topology in our notation, black curves and
    Brazilian bands) and on a signal with two photon jets, each consisting of
    two photons (the $(22)$ topology, red curves). To derive the observed limits,
    we have assumed that the $\eta$ distributions of the data and
    the conversion ratios follow the predictions from Monte Carlo simulations.
    We show results for an
    integrated luminosity of 3.2~fb$^{-1}$, corresponding to the data published
    in~\cite{ATLAS-CONF-2015-081}, and for an integrated luminosity of
    300~fb$^{-1}$.  The Brazilian bands were obtained using the CL$_s$ method
    \cite{Read:2002hq,Mistlberger:2012rs} as implemented in ROOT.
    The blue contours show the discrimination power between the $(11)$
    and $(22)$ scenarios, defined here as the confidence level at which
    the $(11)$ hypothesis can be rejected if the signal in the data consists of
    $(22)$ photon jets.}
  \label{fig:brazilian-bands}
\end{figure}

In \cref{fig:brazilian-bands}, we show the expected and observed limits on such
resonances in the ATLAS diphoton data with 3.2~fb$^{-1}$ of 13~TeV
data~\cite{ATLAS-CONF-2015-081}, and the expected future limits in
300~fb$^{-1}$ of data. Note that the observed limits shown in
  \cref{fig:brazilian-bands} are based on the published ATLAS data,
assuming that the $\eta$ distributions and the conversion rates
(which are not public) follow the predictions from simulations.
Comparing the limits on $(11)$ resonances to those on
$(22)$ resonances, we observe a mild improvement in the latter case.

%--------------------------------------------------------------------%
\section{Long-lived Intermediate States}
\label{sec:finite-tau}
%--------------------------------------------------------------------%

So far, we have assumed that the photon jets in a model with $N_1 > 1$ or $N_2
> 1$ form instantaneously at the primary interaction vertex.  We consider now a
more general scenario, where the intermediate particle $a$ has a non-negligible
proper lifetime $\tau$.  In this scenario, $a$ decays to photons only after
travelling some macroscopic distance $x$ in the inner detector. Since photon
conversion cannot take place until the photons have been produced, the
conversion probability for an individual photon is reduced.  The reduction
factor depends on many parameters, in particular on the distribution of
material in the inner detector and on the efficiency for reconstructing tracks
starting away from the beam axis. A full detector simulation is needed to determine this but a key ingredient is a knowledge of the radial dependence of the conversion probability $p^{\rm conv}$. In the following we outline two simplified approaches.

To obtain an intuitive understanding, it is useful to consider the highly simplistic assumption that the detector is homogeneous. The conversion probability then scales as $1 - x / L_t(\eta)$, where $L_t(\eta)$ is the total distance from the primary vertex to the edge of the tracker.  
%We estimate that this approximation is reasonable for $x \ll
%L_t(\eta)$, while for decay vertices close to $L_t(\eta)$, it probably
%overestimates the tracking efficiency.  
The probability that at least one
photon in an $N$-photon jet converts to an $e^+e^-$ pair inside the tracker is
the probability that $a$ decays between $x-dx$ to $x$, and at least one of the
$N$ photons converts between $x$ and $L_t(\eta)$, integrated over all $x$ from
$0$ to $L_t(\eta)$. This is easy to compute and we find
\begin{align}
  p_N^\text{conv}(\eta, \tau)
    &= \int_0^{L_t(\eta)} \! dx \, \frac{1}{\gamma\tau}
         e^{-x / (\gamma\tau)}
         \bigg[1 - \bigg(1 - p^\text{conv}(\eta) 
                           \bigg(1 - \frac{x}{L_t(\eta)} \bigg) \bigg)^N
         \bigg] \,,
  \label{eq:N-conv-tau}
\end{align}
where $\gamma$ is the Lorentz boost of $a$ and $p^\text{conv}(\eta)$ on the
right hand side is, as in \cref{sec:converted-gammas}, the probability for a
photon to convert between the point of production at the origin and the edge of
the tracker at a distance $L_t(\eta)$.  We have in particular, for $N=1,2$:
\begin{align}
  p_1^\text{conv}(\eta, \tau)
    &= p^\text{conv}(\eta) \left[ 1 - \left(1 - e^{-\frac{L_t(\eta)}{\gamma\tau}} \right)
       \frac{\gamma\tau}{L_t(\eta)} \right] \,,
       \label{eq:1-conv-tau} \\
  p_2^\text{conv}(\eta, \tau)
    &= 2 p^\text{conv}(\eta)
       \left[ 1 - \left(1 - e^{-\frac{L_t(\eta)}{\gamma\tau}} \right)
       \frac{\gamma\tau}{L_t(\eta)} \right]      \nonumber\\
    &\hspace{2cm}
     - [p^\text{conv}(\eta)]^2 \left[ 1 - \frac{2 \gamma\tau}{L_t(\eta)} +
         2 \left(1 - e^{-\frac{L_t(\eta)}{c\gamma\tau}} \right)
         \frac{\gamma^2\tau^2}{L_t^2(\eta)} \right] \,.
       \label{eq:2-conv-tau}
\end{align}
Analogously, the probability for an $N$-photon jet to be detected without any
of the photons converting is
\begin{align}
  p_N^\text{no-conv}(\eta, \tau)
    &= 1 - e^{-L_c(\eta) / (\gamma\tau)} - p_N^\text{conv}(\eta, \tau) \,.
  \label{eq:N-noconv-tau}
\end{align}
Here, the first two terms give the probability that the photon jet is detected
at all, i.e.\ that $a$ decays before reaching the calorimeter at a distance $L_c(\eta)$
from the primary vertex. Note that, because of this factor, $p_N^\text{conv}(\eta, \tau)
+ p_N^\text{no-conv}(\eta, \tau) < 1$.

For obtaining our numerical results, we model the conversion probability density as a function of the radial distance traveled using an approximate ``two-zone'' model based on ref.\,~\cite{ATLAS:2010aba}. In the central region ($|\eta| <0.6$) there are 70\% conversions in the range $(0<r<15\,{\rm cm})$ and 30\% in $(15\,{\rm cm} < r < 40\,{\rm cm})$. In the forward region $(1.3<|\eta|<1.7)$ there are 65\% conversions in $(0<r<15\,{\rm cm})$ and 35\% in $(15\,{\rm cm} < r < 40\,{\rm cm})$, where $r$ is the radial distance. Thus, $1 - x / L_t(\eta)$ is replaced by the above. The total conversion probability remains the same as before.

In an event with two photon jets, the boost factors $\gamma_1$, $\gamma_2$ for the two jets
are in general different. Therefore, in the following numerical analysis, we fold
the conversion probabilities with the distribution of $\gamma_1$, $\gamma_2$
in each $(\eta_1, \eta_2)$ bin, obtained from the same MadGraph simulation that
determines the $(\eta_1, \eta_2)$ distribution (see \cref{sec:converted-gammas}).
Afterwards, the analysis proceeds in the same way as in \cref{sec:converted-gammas}.
In particular, the probability for zero, one or two of the photon jets in an
event to convert are given by \cref{eq:p0,eq:p1,eq:p2}, with the probabilities
$p_N^\text{conv}$ and $p_N^\text{no-conv}$ on the right hand side replaced
by the two-zone analog of \cref{eq:N-conv-tau,eq:N-noconv-tau}. The statistical analysis follows again the procedure described in \cref{sec:statistics}.

\begin{figure}
  \begin{center}
    \begin{tabular}{c}
      \includegraphics[width=11cm]{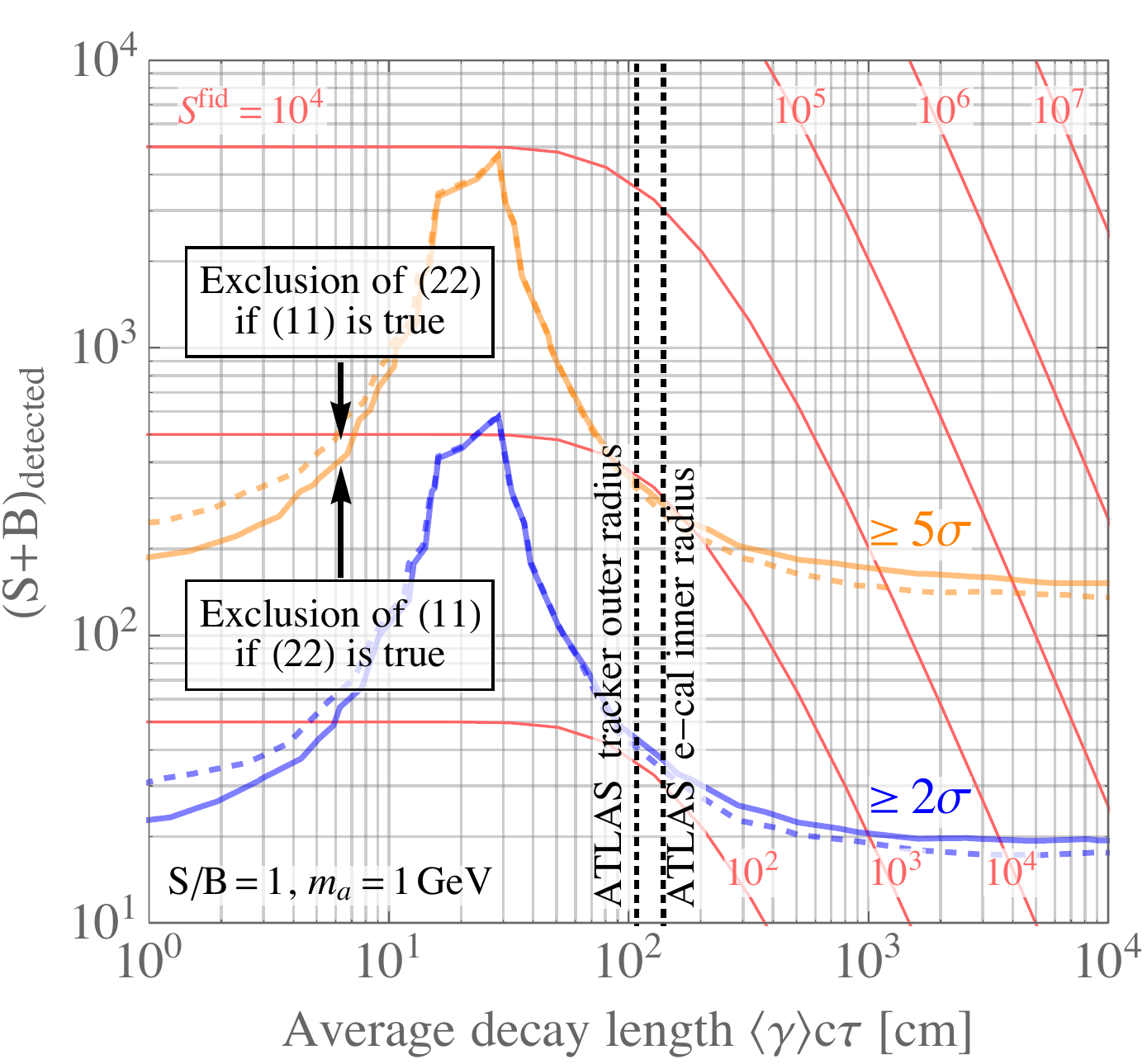}
    \end{tabular}
  \end{center}
  \caption{Number of events required to discriminate between a model
    predicting an $S \to \gamma\gamma$ signal and an alternative model predicting
    an $S \to aa \to 4\gamma$ final state.  Results are shown as a function of
    the lifetime $\tau$ of the intermediate particle, multiplied by their
    average Lorentz boost $\ev{\gamma}$. For definiteness, we assume a
    signal-to-background ratio $S/B = 1$, and an $a$ mass of 1~GeV.
    Note that $(S+B)_{\rm detected}$ only counts those events for which $a$ decays before reaching the EM calorimeter. For $\langle \gamma\rangle c\tau$ larger than the inner e-cal radius, this is only a small fraction of the total number of required events, as indicated by the contours of constant total fiducial signal rate $S^{\rm fid}$. 
    }
\label{fig:sigSBmu}
\end{figure}

In \cref{fig:sigSBmu}, we show the number of expected signal events $S+B$ required
to discriminate between models of $(11)$ and $(22)$ type as a function of the lifetime
of the intermediate particle $a$ in the $(22)$ model.  We take the mass
of the heavy resonance decaying to photon jets to be 750~GeV, the mass of $a$ to be
1~GeV, and we assume the model predicts a signal-to-background ratio $S/B = 1$.
We emphasize that the vertical axis in \cref{fig:sigSBmu} shows the expected number
of \emph{detected} diphoton events. Since for non-negligible $\tau$, only those events
where both $a$ particles decay before entering the calorimeter are detected, we
also show for comparison the \emph{total} number of signal events in the $(22)$
case (red contours in \cref{fig:sigSBmu}).

As is to be expected, the discrimination power is best when $\ev{\gamma}\!\tau =
0$, and worsens for longer lifetimes because a decay away from the beam axis
(but still well within the tracker) leads to a decreased conversion probability
in the $(22)$ model.  When $\ev{\gamma}\!\tau \gtrsim L_c$, it is likely
that $a$ does not decay before reaching the calorimeter, so that events are no
longer categorized as diphoton events. However, among those events which are
detected, the fraction of converted events \emph{increases} again.  Since the
vertical axis in \cref{fig:sigSBmu} shows only the number
$(S+B)_\text{detected}$ of \emph{detected} diphoton events, the discrimination
power based on $(S+B)_\text{detected}$ thus appears to \emph{improve} again in
this case.  Note, however, that the condition $S/B=1$ requires a significantly
larger cross section $\sigma_\text{fid} \, \text{BR}_{\gamma\gamma}$ when
$\ev{\gamma}\!\tau$ is large.

%--------------------------------------------------------------------%
\section{Dark Photons}
\label{sec:dark photons}
%--------------------------------------------------------------------%

An interesting class of models that could in principle mimic a diphoton
resonance signal are those where a new heavy particle $S$ decays to two
\emph{dark photons}---the gauge bosons of a new $U(1)'$ gauge symmetry, often
hypothesized in the context of dark matter models~\cite{Essig:2013lka}.  If the dark
photon $A'$ is sufficiently light ($< m_\mu/2$), its dominant decay mode is $A'
\to e^+ e^-$, so that the process $p p \to S \to (A' \to e^+e^-) + (A' \to
e^+e^-)$ has the same final state as $p p \to S \to \gamma\gamma$, with both
photons converting to $e^+ e^-$ pairs.\footnote{We would like to thank Felix Yu
for pointing out this possibility to us.} While prompt $e^+e^-$ pairs will be vetoed
in the photon reconstruction, if the dark photon lifetime is such that it mostly
decays in the tracker, then these events could easily appear as a diphoton resonance. 

\begin{figure}
  \centering
  \begin{tabular}{c}
    \includegraphics[width=0.5\textwidth]{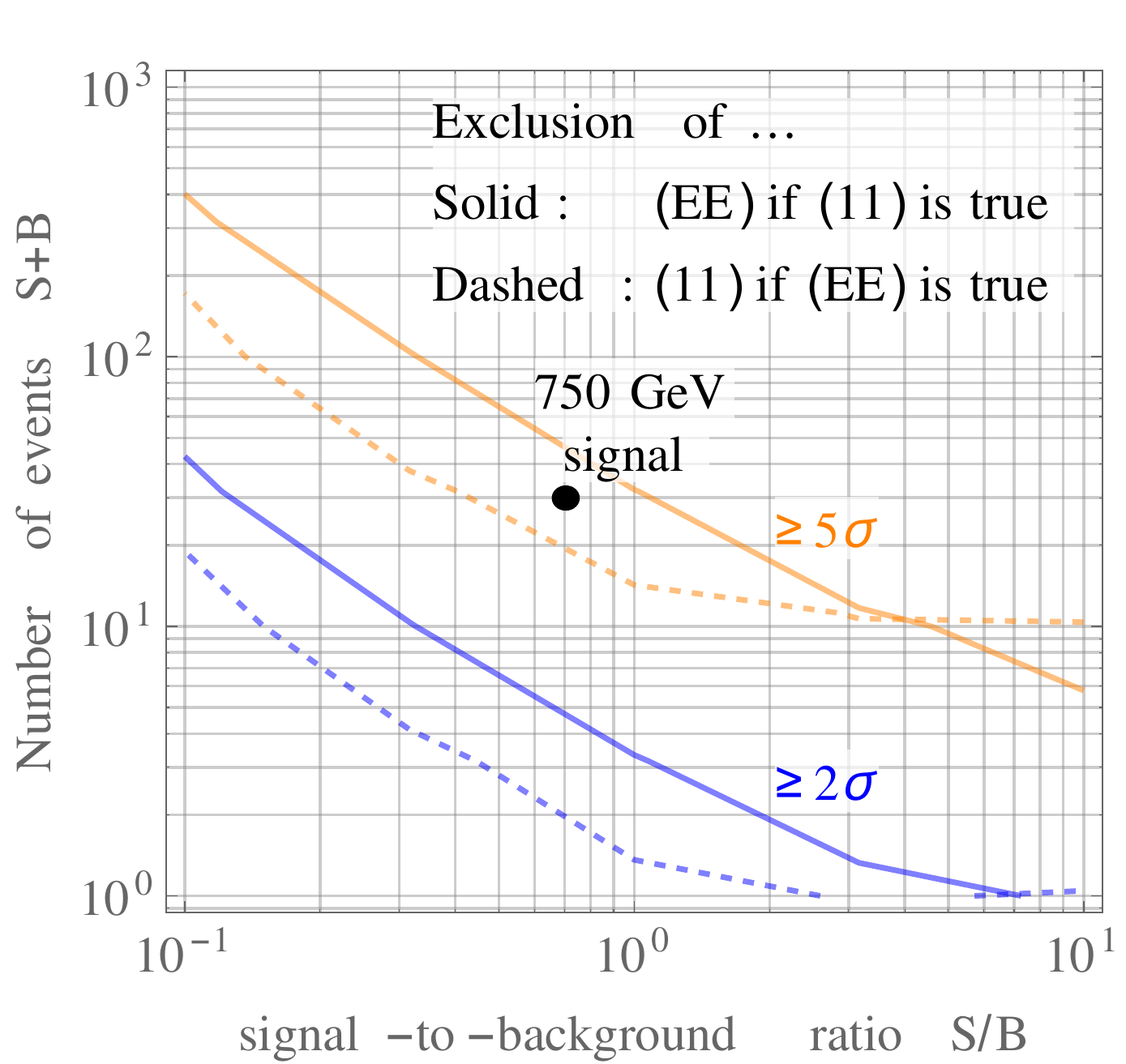}
  \end{tabular}
  \caption{The expected number of diphoton events required to discriminate
    between a true diphoton signal ($(11)$ model) and a signal of the form
    $p p \to S \to (A' \to e^+e^-) + (A' \to e^+e^-)$, where $A'$ is a
    dark photon.  Results are shown as a function of the expected signal-to-background
    ratio.  They are based on a likelihood ratio test, as discussed in
    \cref{sec:statistics}.}
  \label{fig:EE}
\end{figure}

At close inspections, the two topologies are of course different: first
and foremost, $A'$ decays can only mimic converted photons. With good
statistics, it should therefore be easy to tell an $A' A'$ signal apart from true
isolated diphoton signal.  This is indeed the case, as illustrated in \cref{fig:EE}:
at a signal-to-background ratio of $\mathcal{O}(1)$, even a handful of events
is enough to discriminate between the $A'$ model, denoted here as (EE), and
a $(11)$ type diphoton model. Even for a $5\sigma$ test, $\ll 100$~events
are needed.

However, the difference in the apparent photon conversion probabilities
is not the end of the story.  For short-lived $A'$ decaying
quasi-instantaneously, the $e^+e^-$ tracks in the (EE) model will come
directly from the primary vertex, while the tracks from a converted
$\gamma$ can originate at any radius $r$ inside the tracker. Therefore, taking the
radial distribution of the secondary vertices into account, the discrimination
power can be boosted further. We have refrained from doing so in \cref{fig:EE}
to be conservative and because dark photons
could also have a macroscopic decay length. In the latter case, the $r$
distribution predicted by the (EE) model is much more similar to that of a converted $\gamma$
signal in the (11) scenario.  Of course, small differences remain. For instance,
an $A'$ decay can occur anywhere in the tracker, while photon conversion is
only possible inside layers of detector material. Moreover, for $A'$ decay lengths
comparable to the size of the tracker, the $r$ distribution
in the (EE) model is exponentially falling, while in the (11) model it is
constant.

Note that, without inclusion of the $r$ discrimination of the secondary vertices,
the discrimination power depends on the laboratory frame $A'$ decay
length $\gamma\tau$ only through the factor
$[1 - \exp(L_c(\eta_1)/(\gamma\tau)] \, [1 - \exp(L_c(\eta_2)/(\gamma\tau)]$,
which gives the probability that both $A'$ decays occur before the calorimeter
(see \cref{sec:finite-tau}). In other words, if $S/B$ and
$(S+B)_\text{detected}$ are fixed, as in \cref{fig:sigSBmu}, the discrimination
power is independent of $\ev{\gamma}\!\tau$.

%--------------------------------------------------------------------%
\section{Further Observables}
\label{sec:other-observables}
%--------------------------------------------------------------------%

Photon reconstruction in the LHC detectors offers additional handles that could
be used to further discriminate photon jets from isolated photons, and possibly
pin down underlying structures like the multiplicity of photons inside the jet.
In the following we briefly discuss the most promising ideas:
\begin{itemize}
  \item The photon pairs coming from a boosted decay $a\to \gamma\gamma$ carry
    roughly equal energy. If only one of them converts, the ratio of track
    $p_T$ to calorimeter energy $E_{\rm cal}$ should differ substantially
    from one, the value expected for single photons.
    This is a very powerful variable that is not
    currently being used. 
    An accurate measurement of the electron track $p_T$ is complicated by their relatively 
    large $p_T\sim 100$~GeV and the fact that they only traverse part of the tracker, depending
    on where they convert. Therefore a sufficient number of events is needed such that 
    the measurement can be made on those events where both electrons from the conversion
    are well reconstructed. 
    
  \item When more than one photon inside a photon jets converts to $e^+e^-$,
    up to $2N$ tracks could be reconstructed for an $N$-photon jet. Such multiple
    conversions might be rejected by the standard photon reconstruction algorithms, 
    for example in ATLAS~\cite{ATLAS:2011kuc} a cut is placed on the $p_T$ sum of 
    tracks within $\Delta R=0.3$ of the photon candidate which are not associated with
    the photon candidate itself. 
    Therefore we expect that the reconstruction efficiency for $N>1$ photon jets is reduced.
    However once a resonance is found, events with larger track multiplicities
    can be explicitly searched for in loose photon samples to get additional
    information on the signal. 

  \item Photon jet events will react differently from single photons to changes
    in the isolation criteria. While variables which mainly cut on nearby
    hadronic activity are insensitive to the photon multiplicity, those using
    electromagnetic calorimeter shower shapes could be very sensitive.
    In the ATLAS search for Higgs decays to pairs of photon jets~\cite{ATLAS:2012soa}, it was
    already shown that the variable $F_{\rm side}$,
    which considers the ratio of energies deposited in 3 vs.\ 7 bins centered on
    the highest bin, is very sensitive to the mass of the intermediate
    pseudo-scalar $m_a$. A large change in efficiency of $F_{\rm side}$ in the
    signal region compared to the side-bands would therefore be a strong
    indication of photon jets, and even give direct access to $m_a$.

    In this context, it is worth commenting on photon jets with $N>2$ constituent
    photons.  At first glance it seems unlikely that such final states could
    successfully mimic an isolated photon signal, given how difficult it already
    is to sufficiently collimate two photons. However, one should also note that
    for $N>2$, the energy does not have to be distributed evenly between the photons.
    For instance, if one is significantly harder than the others, the energy deposit
    in the electromagnetic calorimeter would have a single peak structure such that
    rejection methods based on the shape of the calorimeter cluster would fail.
\end{itemize}
Each of these strategies can provide additional insight into the nature of a
diphoton signal which might be discovered in the future, or more general into
any new physics signals involving photons. Comparison with control regions and
side bands can be used to verify that abnormal behavior of the photon
candidates in the signal region is indeed due to photon jets and not just from
e.g.\ QCD backgrounds.

%--------------------------------------------------------------------%
\section{Conclusions}
\label{sec:conclusions}
%--------------------------------------------------------------------%

To summarize, we have discussed from a phenomenologist's point of view how
the conversion of photons to $e^+ e^-$ pairs inside the LHC detectors can
be exploited to discriminate between final states involving isolated photons
and events containing jets of multiple highly collimated photons. Such
photon jets arise, for instance, when a light new particle is produced on-shell
and decays to two photons.  We have illustrated that, even with modest statistics,
a resonance decaying to two isolated photons can be distinguished
from a new particle decaying to two photon jets.

For instance, in the context of the possible 750~GeV resonance observed
in ATLAS and CMS data, $\sim 30$ events, are sufficient to make this
distinction at the $2\sigma$ level, while $\mathcal{O}(100)$ events
are required for a $5\sigma$ discrimination.

We have also illustrated how the sensitivity to photon jet signals
mimicking a diphoton resonance depends mildly on the multiplicity of the photon jets.
Finally, we have studied scenarios in which photon jets emerge
at a macroscopic distance from the beam pipe in the
decay of a long-lived intermediate particle.

We conclude that photon candidates in the LHC detectors offer an extremely rich
substructure which can be exploited for highly efficient model discrimination.
This substructure is theoretically well modeled and seems readily
accessible experimentally. We hope that the results presented in this note
will be useful in this endeavor.

%--------------------------------------------------------------------%
\section*{Acknowledgments}
%--------------------------------------------------------------------%

We would like to thank F.~Kahlhoefer, A.~Katz, K.~Schmidt-Hoberg,
and F.~Yu for useful discussions, and especially K.~Peters for useful discussions and
many helpful comments on the draft. 
The work of JK is supported by the German Research Foundation (DFG)
in the framework of the Research Unit ``New Physics at the Large Hadron
Collider'' (FOR~2239) and of Grant No.\ \mbox{KO~4820/1--1}, and by
the European Research Council (ERC) under the European Union's
Horizon 2020 research and innovation programme (grant agreement No.\ 637506,
``$\nu$Directions'').  Additional support has been provided by the Cluster of
Excellence ``Precision Physics, Fundamental Interactions and Structure of
Matter'' (PRISMA -- EXC 1098), grant No.~05H12UME of the German Federal
Ministry for Education and Research (BMBF).

%--------------------------------------------------------------------%
\appendix
\section{Background Dependence}
\label{sec:backgrnds}
%--------------------------------------------------------------------%
The diphoton background usually has three components, which besides pairs of prompt photons includes events where either one or both photons are misidentified jets which essentially are due to neutral hadrons decaying to photon pairs. Thus the background too can have events of $(1 2)$ and $(2 2)$ type.
  
In the 750~GeV signal window, the jet contributions to the diphoton backgrounds are of order 10\% in CMS~\cite{CMS-PAS-EXO-15-004}, and probably of the same order in ATLAS (see e.g. Fig. 7 in the supplemental material for Ref.~\cite{Aad:2014ioa}). 
Nevertheless we would like to stress that our method also works for different background composition, which could be relevant for applications in other search channels. 

\begin{figure}[!b]
\includegraphics[width=0.5\textwidth]{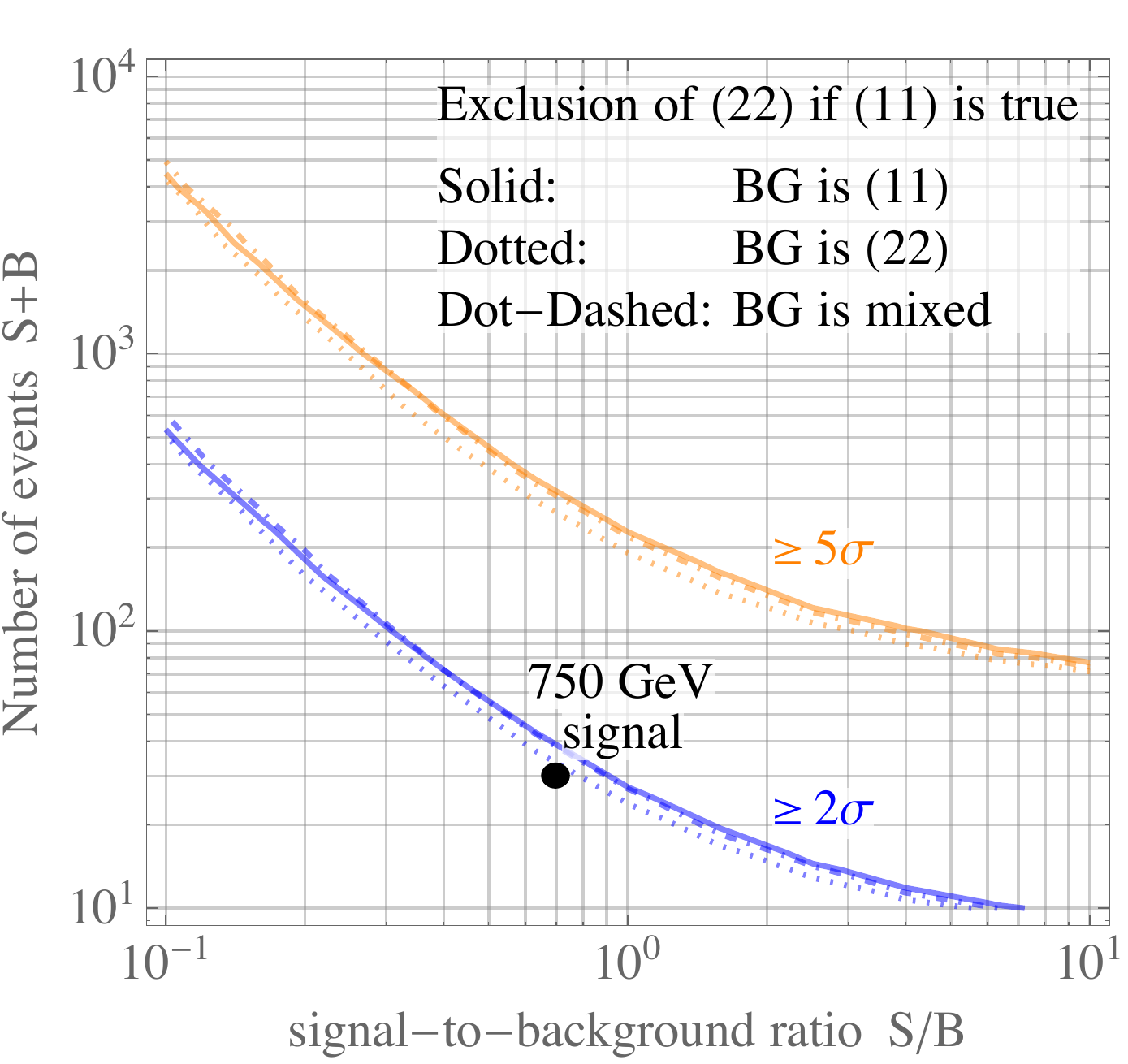}
\caption{Discrimination power for different background compositions. For the mixed case we assume that half the background is of $(1 1)$ type and the other half of $(2 2)$ type. }
\label{fig:bgvariations}
\end{figure}

In Fig.~\ref{fig:bgvariations} we see that the discrimination power is only marginally affected if the background composition is varied. Therefore as long as the background composition can be measured in control regions, the discrimination power will remain.

%--------------------------------------------------------------------%
\section{Statistical Procedure}
\label{sec:statistics}
%--------------------------------------------------------------------%
Different theoretical models leading to a diphoton-like signature can
be distinguished using a likelihood ratio test (see e.g.\ \cite{Agashe:2014kda}).
The test statistic is
\begin{align}
  \text{LLR}(D, M_1, M_2) = \log\bigg( \frac{\mathcal{L}(D | M_1)}
                                            {\mathcal{L}(D | M_2)} \bigg) \,,
  \label{eq:LLR}
\end{align}
where $\mathcal{L}(\text{data} | M_m)$ denotes the likelihood of the data $D$
if the model hypothesis $M_m$ is true.  A model hypothesis is characterized
here by the multiplicities $(N_1 N_2)$ of the two photon jets in each signal
event, and by the associated prediction $P_{M_m}^i$ for the event rate
in the $i$-th bin.  The likelihood is given by
\begin{align}
  \log \mathcal{L}(D | M_m) =
    \sum_i \big[
        2 ( D^i - P_{M_m}^i)
      + 2 D^i \log \frac{D^i}{P_{M_m}^i}
  \big] \,.
  \label{eq:likelihood-Poisson}
\end{align}
In the simplest case, the bin index $i=0,1,2$ denotes the number of photon jets
in the event that are reconstructed as converted photons.  However, since the
conversion probability depends on the rapidities $\eta_1$, $\eta_2$ of the
photon jets and since the rapidity dependence is different for signal and
background events, we bin the data also in $|\eta_1|$, $|\eta_2|$.  This turns
$i$ into a multi-index $(i j k)$, with $i = 0,1,2$
the number of converted photon jets, and $j$, $k$ denoting
the rapidity bins.  We use four of the latter for each photon jet, as given in
\cref{tab:eta-binning}.  

\begin{table}[!h]
  \centering
  \begin{minipage}{7cm}
    \begin{ruledtabular}
      \begin{tabular}{lcc}
        \# & bin boundaries in $\eta$ & $p^\text{conv}(\eta)$ \\
        \hline
         1 & $[0,    0.59]$ & 0.25                  \\
         2 & $[0.59, 1.35]$ & 0.39                  \\
         3 & $[1.52, 1.78]$ & 0.54                  \\
         4 & $[1.78, 2.4]$  & 0.47                  \\
      \end{tabular}
    \end{ruledtabular}
  \end{minipage}
  \caption{Rapidity bins used in our analysis, together with the associated
    probabilities $p^\text{conv}$ for a single photon in a given bin to
    convert into an $e^+e^-$ pair.}
  \label{tab:eta-binning}
\end{table}

Note that this binning excludes the transition region
between the barrel and the endcap. In the notation of \cref{sec:converted-gammas},
the number of predicted events $P_{M_m}^{(ijk)}$
for a model $M_m = (N_1 N_2)$ is given by
\begin{align}
  P_{M_m}^{(ijk)}
    = B \, p_B^{jk} \, p_{(11)}^{(i)}(\eta_1^j, \eta_2^k)
    + S \, p_S^{jk} \, p_{(N_1 N_2)}^{(i)}(\eta_1^j, \eta_2^k) \,.
\end{align}

We wish to compute the expected confidence level at which model $M_2$ can be
ruled out in favor of $M_1$, if $M_1$ is realized in nature. To do so, we
generate $\mathcal{O}(10^4)$ sets $\{ \tilde{D}^{(ijk)} \}$ of pseudo-data
distributed according to model $M_2$ and compute the log-likelihood ratio
$\text{LLR}(\tilde{D}, M_1, M_2)$ for each of them. We thus obtain the
probability distribution function (PDF) of $\text{LLR}(\tilde{D}, M_1, M_2)$. We then
compute also the log-likelihood ratio $\text{LLR}(M_1, M_1, M_2)$ for the case
that the data $D^{(ijk)}$ equals the prediction $P_{M_1}^{ijk}$ of the assumed ``true''
model. Evaluating the cumulative distribution function (CDF) of
$\text{LLR}(\tilde{D}, M_1, M_2)$ at the value $\text{LLR}(M_1, M_1, M_2)$
yields the desired confidence level for the exclusion of $M_2$.

While this Monte Carlo-based method for evaluating confidence intervals is very
general and, by the Neyman--Pearson lemma, offers optimal discrimination
power, it could be replaced by a much simpler $\chi^2$ test. Namely, note that
$-2 \log{\mathcal{L}(D | M_2)}$ follows a $\chi^2$ distribution if the total
number of events predicted by $M_2$ is not too small, The number of degrees of freedom
of the $\chi^2$ distribution is given by the number of bins.  The expected
confidence level at which model $M_2$ is disfavored if $M_1$ is true is thus
given by the CDF of the $\chi^2$ distribution, evaluated at $-2
\log{\mathcal{L}(M_1 | M_2)}$. If the number of events in each bin
is $\gtrsim 10$, so that the Poissonian likelihood \cref{eq:likelihood-Poisson}
is well approximated by the Gaussian likelihood
\begin{align}
  -2 \log \mathcal{L}^\text{Gauss}(D | M_m) =
    \sum_{i=0}^2 \sum_k \frac{(D^{(ijk)} - P_{M_m}^{(ijk)})^2}{P_{M_m}^{(ijk)}} \,,
  \label{eq:likelihood-Gauss}
\end{align}
we recover the $\chi^2$ from \cref{eq:chi2-eta}.

\bibliographystyle{JHEP}
%\interlinepenalty=10000
%\tolerance=100
\bibliography{./photon-conversion}

\providecommand{\href}[2]{#2}\begingroup\raggedright\begin{thebibliography}{10}

\bibitem{ATLAS-CONF-2015-081}
{\bf ATLAS} Collaboration, {\it {Search for resonances decaying to photon pairs
  in 3.2 fb$^{-1}$ of $pp$ collisions at $\sqrt{s}$ = 13 TeV with the ATLAS
  detector}},  Tech. Rep. ATLAS-CONF-2015-081, CERN, Geneva, Dec, 2015.

\bibitem{CMS-PAS-EXO-15-004}
{\bf CMS} Collaboration, {\it {Search for new physics in high mass diphoton
  events in proton-proton collisions at 13TeV}},  Tech. Rep.
  CMS-PAS-EXO-15-004, CERN, Geneva, 2015.

\bibitem{DiphotonInspire}
{The global hep-ph community}, 2016.
\newblock \url{http://inspirehep.net/search?p=refersto%3Arecid%3A1410174}.

\bibitem{Draper:2012xt}
P.~Draper and D.~McKeen, {\it {Diphotons from Tetraphotons in the Decay of a
  125 GeV Higgs at the LHC}},  {\em Phys. Rev.} {\bf D85} (2012) 115023,
  [\href{http://arxiv.org/abs/1204.1061}{{\tt arXiv:1204.1061}}].

\bibitem{Ellis:2012zp}
S.~D. Ellis, T.~S. Roy, and J.~Scholtz, {\it {Phenomenology of Photon-Jets}},
  {\em Phys. Rev.} {\bf D87} (2013), no.~1 014015,
  [\href{http://arxiv.org/abs/1210.3657}{{\tt arXiv:1210.3657}}].

\bibitem{Ellis:2012sd}
S.~D. Ellis, T.~S. Roy, and J.~Scholtz, {\it {Jets and Photons}},  {\em Phys.
  Rev. Lett.} {\bf 110} (2013), no.~12 122003,
  [\href{http://arxiv.org/abs/1210.1855}{{\tt arXiv:1210.1855}}].

\bibitem{Dobrescu:2000jt}
B.~A. Dobrescu, G.~L. Landsberg, and K.~T. Matchev, {\it {Higgs boson decays to
  CP odd scalars at the Tevatron and beyond}},  {\em Phys. Rev.} {\bf D63}
  (2001) 075003, [\href{http://arxiv.org/abs/hep-ph/0005308}{{\tt
  hep-ph/0005308}}].

\bibitem{Toro:2012sv}
N.~Toro and I.~Yavin, {\it {Multiphotons and photon jets from new heavy vector
  bosons}},  {\em Phys. Rev.} {\bf D86} (2012) 055005,
  [\href{http://arxiv.org/abs/1202.6377}{{\tt arXiv:1202.6377}}].

\bibitem{Chang:2006bw}
S.~Chang, P.~J. Fox, and N.~Weiner, {\it {Visible Cascade Higgs Decays to Four
  Photons at Hadron Colliders}},  {\em Phys. Rev. Lett.} {\bf 98} (2007)
  111802, [\href{http://arxiv.org/abs/hep-ph/0608310}{{\tt hep-ph/0608310}}].

\bibitem{Curtin:2013fra}
D.~Curtin et~al., {\it {Exotic decays of the 125 GeV Higgs boson}},  {\em Phys.
  Rev.} {\bf D90} (2014), no.~7 075004,
  [\href{http://arxiv.org/abs/1312.4992}{{\tt arXiv:1312.4992}}].

\bibitem{Knapen:2015dap}
S.~Knapen, T.~Melia, M.~Papucci, and K.~Zurek, {\it {Rays of light from the
  LHC}},  \href{http://arxiv.org/abs/1512.04928}{{\tt arXiv:1512.04928}}.

\bibitem{Agrawal:2015dbf}
P.~Agrawal, J.~Fan, B.~Heidenreich, M.~Reece, and M.~Strassler, {\it
  {Experimental Considerations Motivated by the Diphoton Excess at the LHC}},
  \href{http://arxiv.org/abs/1512.05775}{{\tt arXiv:1512.05775}}.

\bibitem{Chala:2015cev}
M.~Chala, M.~Duerr, F.~Kahlhoefer, and K.~Schmidt-Hoberg, {\it {Tricking
  Landau-Yang: How to obtain the diphoton excess from a vector resonance}},
  \href{http://arxiv.org/abs/1512.06833}{{\tt arXiv:1512.06833}}.

\bibitem{Aparicio:2016iwr}
L.~Aparicio, A.~Azatov, E.~Hardy, and A.~Romanino, {\it {Diphotons from
  Diaxions}},  \href{http://arxiv.org/abs/1602.00949}{{\tt arXiv:1602.00949}}.

\bibitem{Chang:2015sdy}
J.~Chang, K.~Cheung, and C.-T. Lu, {\it {Interpreting the 750 GeV Di-photon
  Resonance using photon-jets in Hidden-Valley-like models}},
  \href{http://arxiv.org/abs/1512.06671}{{\tt arXiv:1512.06671}}.

\bibitem{Ellwanger:2016qax}
U.~Ellwanger and C.~Hugonie, {\it {A 750 GeV Diphoton Signal from a Very Light
  Pseudoscalar in the NMSSM}},  \href{http://arxiv.org/abs/1602.03344}{{\tt
  arXiv:1602.03344}}.

\bibitem{ATLAS-gamma-conv}
{\it {Photon conversion reconstruction}},  Tech. Rep. ATLAS-EGAM-2015-004,
  CERN, Geneva, Dec, 2015.

\bibitem{Khachatryan:2015iwa}
{\bf CMS} Collaboration, V.~Khachatryan et~al., {\it {Performance of Photon
  Reconstruction and Identification with the CMS Detector in Proton-Proton
  Collisions at sqrt(s) = 8 TeV}},  {\em JINST} {\bf 10} (2015), no.~08 P08010,
  [\href{http://arxiv.org/abs/1502.02702}{{\tt arXiv:1502.02702}}].

\bibitem{Bjorken:2009mm}
J.~D. Bjorken, R.~Essig, P.~Schuster, and N.~Toro, {\it {New Fixed-Target
  Experiments to Search for Dark Gauge Forces}},  {\em Phys. Rev.} {\bf D80}
  (2009) 075018, [\href{http://arxiv.org/abs/0906.0580}{{\tt
  arXiv:0906.0580}}].

\bibitem{Jaeckel:2010ni}
J.~Jaeckel and A.~Ringwald, {\it {The Low-Energy Frontier of Particle
  Physics}},  {\em Ann.Rev.Nucl.Part.Sci.} {\bf 60} (2010) 405--437,
  [\href{http://arxiv.org/abs/1002.0329}{{\tt arXiv:1002.0329}}].

\bibitem{Essig:2010gu}
R.~Essig, R.~Harnik, J.~Kaplan, and N.~Toro, {\it {Discovering New Light States
  at Neutrino Experiments}},  {\em Phys.Rev.} {\bf D82} (2010) 113008,
  [\href{http://arxiv.org/abs/1008.0636}{{\tt arXiv:1008.0636}}].

\bibitem{Essig:2013lka}
R.~Essig et~al., {\it {Working Group Report: New Light Weakly Coupled
  Particles}},  in {\em {Community Summer Study 2013: Snowmass on the
  Mississippi (CSS2013) Minneapolis, MN, USA, July 29-August 6, 2013}}, 2013.
\newblock \href{http://arxiv.org/abs/1311.0029}{{\tt arXiv:1311.0029}}.

\bibitem{Alekhin:2015byh}
S.~Alekhin et~al., {\it {A facility to Search for Hidden Particles at the CERN
  SPS: the SHiP physics case}},  \href{http://arxiv.org/abs/1504.04855}{{\tt
  arXiv:1504.04855}}.

\bibitem{Jaeckel:2015jla}
J.~Jaeckel and M.~Spannowsky, {\it {Probing MeV to 90 GeV axion-like particles
  with LEP and LHC}},  {\em Phys. Lett.} {\bf B753} (2016) 482--487,
  [\href{http://arxiv.org/abs/1509.00476}{{\tt arXiv:1509.00476}}].

\bibitem{Dobrich:2015jyk}
B.~Döbrich, J.~Jaeckel, F.~Kahlhoefer, A.~Ringwald, and K.~Schmidt-Hoberg, {\it
  {ALPtraum: ALP production in proton beam dump experiments}},  {\em JHEP} {\bf
  02} (2016) 018, [\href{http://arxiv.org/abs/1512.03069}{{\tt
  arXiv:1512.03069}}]. [JHEP02,018(2016)].

\bibitem{Aad:2015asa}
{\bf ATLAS} Collaboration, G.~Aad et~al., {\it {Search for pair-produced
  long-lived neutral particles decaying in the ATLAS hadronic calorimeter in
  $pp$ collisions at $\sqrt{s}$ = 8 TeV}},  {\em Phys. Lett.} {\bf B743} (2015)
  15--34, [\href{http://arxiv.org/abs/1501.04020}{{\tt arXiv:1501.04020}}].

\bibitem{Aad:2015rba}
{\bf ATLAS} Collaboration, G.~Aad et~al., {\it {Search for massive, long-lived
  particles using multitrack displaced vertices or displaced lepton pairs in pp
  collisions at $\sqrt{s}$ = 8 TeV with the ATLAS detector}},  {\em Phys. Rev.}
  {\bf D92} (2015), no.~7 072004, [\href{http://arxiv.org/abs/1504.05162}{{\tt
  arXiv:1504.05162}}].

\bibitem{Alwall:2011uj}
J.~Alwall, M.~Herquet, F.~Maltoni, O.~Mattelaer, and T.~Stelzer, {\it {MadGraph
  5 : Going Beyond}},  {\em JHEP} {\bf 1106} (2011) 128,
  [\href{http://arxiv.org/abs/1106.0522}{{\tt arXiv:1106.0522}}].

\bibitem{Alwall:2014hca}
J.~Alwall, R.~Frederix, S.~Frixione, V.~Hirschi, F.~Maltoni, et~al., {\it {The
  automated computation of tree-level and next-to-leading order differential
  cross sections, and their matching to parton shower simulations}},
  \href{http://arxiv.org/abs/1405.0301}{{\tt arXiv:1405.0301}}.

\bibitem{Alloul:2013bka}
A.~Alloul, N.~D. Christensen, C.~Degrande, C.~Duhr, and B.~Fuks, {\it
  {FeynRules 2.0 - A complete toolbox for tree-level phenomenology}},
  \href{http://arxiv.org/abs/1310.1921}{{\tt arXiv:1310.1921}}.

\bibitem{Read:2002hq}
A.~L. Read, {\it {Presentation of search results: The CL(s) technique}},  {\em
  J.Phys.} {\bf G28} (2002) 2693--2704.

\bibitem{Mistlberger:2012rs}
B.~Mistlberger and F.~Dulat, {\it {Limit setting procedures and theoretical
  uncertainties in Higgs boson searches}},
  \href{http://arxiv.org/abs/1204.3851}{{\tt arXiv:1204.3851}}.

\bibitem{ATLAS:2010aba}
{\bf ATLAS} Collaboration, {\it {Photon Conversions at sqrt{s} = 900 GeV
  measured with the ATLAS Detector}}, .

\bibitem{ATLAS:2011kuc}
{\bf ATLAS} Collaboration, {\it {Expected photon performance in the ATLAS
  experiment}},  Tech. Rep. ATL-PHYS-PUB-2011-007, ATL-COM-PHYS-2010-1051,
  2011.

\bibitem{ATLAS:2012soa}
{\bf ATLAS} Collaboration, {\it {Search for a Higgs boson decaying to four
  photons through light CP-odd scalar coupling using 4.0~fb$^{-1}$ of
  $7~\mathrm{TeV}$ $pp$ collision data taken with ATLAS detector at the LHC}},
  Tech. Rep. ATLAS-CONF-2012-079, 2012.

\bibitem{Aad:2014ioa}
{\bf ATLAS} Collaboration, G.~Aad et~al., {\it {Search for Scalar Diphoton
  Resonances in the Mass Range $65-600$ GeV with the ATLAS Detector in $pp$
  Collision Data at $\sqrt{s}$ = 8 $TeV$}},  {\em Phys. Rev. Lett.} {\bf 113}
  (2014), no.~17 171801, [\href{http://arxiv.org/abs/1407.6583}{{\tt
  arXiv:1407.6583}}].

\bibitem{Agashe:2014kda}
{\bf Particle Data Group} Collaboration, K.~A. Olive et~al., {\it {Review of
  Particle Physics}},  {\em Chin. Phys.} {\bf C38} (2014) 090001.

\end{thebibliography}\endgroup

\end{document}